\documentclass{article}

\usepackage{microtype}
\usepackage{graphicx}
\usepackage{booktabs} 
\usepackage{amsmath}
\usepackage{amssymb}
\usepackage[colorinlistoftodos,textsize=scriptsize]{todonotes}
\usepackage{marginnote}
\usepackage{amsthm}
\usepackage{booktabs}
\usepackage{multirow}
\usepackage{float}
\usepackage{subcaption}
\usepackage{wrapfig}
\graphicspath{{./Figures/}}
\usepackage{hyperref}

\usepackage{amsmath,amsfonts,bm}

\def\eqref#1{equation~\ref{#1}}

\def\1{\bm{1}}

\DeclareMathAlphabet{\mathsfit}{\encodingdefault}{\sfdefault}{m}{sl}
\SetMathAlphabet{\mathsfit}{bold}{\encodingdefault}{\sfdefault}{bx}{n}

\usepackage{hyperref}

\usepackage[accepted]{icml2021}

\icmltitlerunning{\ours: Formula Prediction from Semi-structured Context}

\begin{document}
\newcommand{\eat}[1]{}
\newtheorem{defi}{Definition}

\newif\ifsubmit
\submittrue

\newcommand{\maybetodo}[1]{\ifsubmit{}\else{#1}\fi}

\newcommand{\ours}{\textsc{SpreadsheetCoder}}

\newcommand{\xinyun}[1]{{\maybetodo{\color{blue}[Xinyun: #1]}}}
\newcommand{\hadai}[1]{{\maybetodo{\color{magenta} [Hanjun: #1]}}}
\newcommand{\rishabh}[1]{{\maybetodo{\color{green} [Rishabh: #1]}}}
\newcommand{\maniatis}[1]{{\maybetodo{\color{pink} [Petros: #1]}}}
\newcommand{\charles}[1]{{\maybetodo{\color{Plum} [Charles: #1]}}}

\twocolumn[
\icmltitle{\ours: Formula Prediction from Semi-structured Context}

\begin{icmlauthorlist}
\icmlauthor{Xinyun Chen}{ucb}
\icmlauthor{Petros Maniatis}{google}
\icmlauthor{Rishabh Singh}{google}
\icmlauthor{Charles Sutton}{google}
\icmlauthor{Hanjun Dai}{google}
\icmlauthor{Max Lin}{google}
\icmlauthor{Denny Zhou}{google}
\end{icmlauthorlist}

\icmlaffiliation{ucb}{UC Berkeley}
\icmlaffiliation{google}{Google}

\icmlcorrespondingauthor{Xinyun Chen}{xinyun.chen@berkeley.edu}
\icmlkeywords{Neural program synthesis, Spreadsheet formula prediction}

\vskip 0.3in]



\printAffiliationsAndNotice{}  

\begin{abstract}
Spreadsheet formula prediction has been an important program synthesis problem with many real-world applications. Previous works typically utilize input-output examples as the specification for spreadsheet formula synthesis, where each input-output pair simulates a separate row in the spreadsheet. However, this formulation does not fully capture the rich context in real-world spreadsheets. First, spreadsheet data entries are organized as tables, thus rows and columns are not necessarily independent from each other. In addition, many spreadsheet tables include headers, which provide high-level descriptions of the cell data. However, previous synthesis approaches do not consider headers as part of the specification. In this work, we present the first approach for synthesizing spreadsheet formulas from tabular context, which includes both headers and semi-structured tabular data. In particular, we propose~\ours, a BERT-based model architecture to represent the tabular context in both row-based and column-based formats. We train our model on a large dataset of spreadsheets, and demonstrate that{~\ours} achieves top-1 prediction accuracy of $42.51\%$, which is a considerable improvement over baselines that do not employ rich tabular context. Compared to the rule-based system, \ours{} assists 82\% more users in composing formulas on Google Sheets.
\end{abstract}

\section{Introduction}
\label{sec:intro}

Spreadsheets are ubiquitous for data storage, with hundreds of millions of users. Helping users write formulas in spreadsheets is a powerful feature for data analysis. Although spreadsheet formula languages are relatively simpler than general-purpose programming languages for data manipulation, writing spreadsheet formulas could still be tedious and error-prone for end users~\citep{gulwani2011automating,hermans2012measuring,cheung2016custodes}. Systems such as FlashFill~\citep{gulwani2011automating,gulwani2012spreadsheet} help end-users perform string transformation tasks in spreadsheets using a few input-output examples by automatically synthesizing a program in a domain-specific language (DSL). Recently, several learning approaches based on different neural architectures have been developed for learning such programs from examples, and have demonstrated promising results~\citep{parisotto2016neuro,devlin2017robustfill,vijayakumar2018neural}.

All these previous works formalize the spreadsheet program prediction problem as a \emph{programming by example} task, with the goal of synthesizing programs from a small number of input-output examples.
We argue that this choice engenders three key limitations.
First, this setup assumes that each data row is independent, and each formula is executed on data cells of the same row. However, real spreadsheets are less structured than this. 
Data in spreadsheets is typically organized as semi-structured tables, and cells in different rows could be correlated. As shown in Figure~\ref{fig:sheets-ex}, in the same table, different data blocks could have different structures, and formulas can take cell values in other rows as function arguments. 
Second, because spreadsheets are semi-structured, they also contain rich metadata. In particular, many spreadsheet tables include headers that provide high-level descriptions of the data, which could provide important clues for formula prediction. However, table headers are not utilized in prior work.
Finally, programming-by-example methods output programs in a 
DSL, which is typically designed to facilitate synthesis, and is much less flexible than the language in which users write formulas.
For example, the FlashFill DSL only covers a subset of spreadsheet functions for string processing, and it does not support rectangular ranges, a common feature of spreadsheet formulas. In contrast, spreadsheet languages also support a wide variety of functions for numerical calculation, while the argument selection is more flexible and takes the spreadsheet table structure into account.
In total, these limitations can compromise the applicability
of such prior efforts to more diverse real-world spreadsheets and to richer language functionality.

Instead, we propose synthesizing spreadsheet formulas \emph{without} an explicit specification. To predict a formula
in a given cell, the context of data and metadata
is used as an \emph{implicit} (partial) specification of the desired program. For example (Figure~\ref{fig:sheets-ex-col}), if predicting a formula at the end of a column of numbers labeled ``Score'', and a cell in the same row contains the text ``Total'',
this context might specify the user's
intent to compute a column sum.
Our problem brings several new challenges  compared to related work in programming by example~\citep{gulwani2011automating,bunel2018leveraging,balog2016deepcoder},
semantic parsing~\citep{popescu2003towards,zhong2017seq2sql,yu2018spider} and source code completion~\citep{raychev2014code,li2018code,svyatkovskiy2019pythia}.
Spreadsheet tables contain rich two-dimensional relational
structure and natural language metadata,
but the rows do not follow a fixed schema
as in a relational database. 
Meanwhile, our tabular context is more ambiguous as the program specification, and the spreadsheet language studied in this work is more flexible
than languages studied in the program synthesis literature.

In this paper, we present~\ours, a neural network architecture for spreadsheet formula prediction. {\ours} encodes the spreadsheet context in its table format, and generates the corresponding formula in the target cell. A BERT-based encoder~\citep{devlin2019bert} computes an embedding vector for each input token,  incorporating the contextual information from  nearby rows and columns. The BERT encoder is initialized from the weights pre-trained on English text corpora, which is beneficial for encoding table headers. 
To handle cell references, we propose a two-stage decoding process inspired by sketch learning for program synthesis~\citep{solar2008program,murali2018neural,dong2018coarse,nye2019learning}. Our decoder first generates a formula sketch, which does not include concrete cell references, and then predicts the corresponding cell ranges to generate the complete formula.

For evaluation (Section~\ref{sec:exp}), we construct a large-scale benchmark of spreadsheets publicly shared within our organization.
We show that $\ours$
outperforms neural network approaches
for programming by example~\citep{devlin2017robustfill}, and achieves $42.51\%$ top-1 full-formula accuracy, and $57.41\%$ top-1 formula-sketch accuracy, both of which are already high enough to be practically useful.
In particular, \ours{} assists 82\% more users in composing formulas than the rule-based system on Google Sheets.
Moreover, {\ours} can predict cell ranges and around a hundred different spreadsheet operators, which is much more flexible
than DSLs used in prior works.
With various ablation experiments, we demonstrate that both implicit specification from the context and
text from the headers are crucial for obtaining good performance.

\vspace{-1em}
\section{Problem Setup}
\label{sec:setup}

In this section, we discuss the setup of our spreadsheet formula prediction problem. We first describe the input specification, then introduce the language and representation for spreadsheet formulas.

\begin{figure*}[t]
    \centering
    \begin{subfigure}[t]{0.45\linewidth}
    \includegraphics[width=\linewidth]{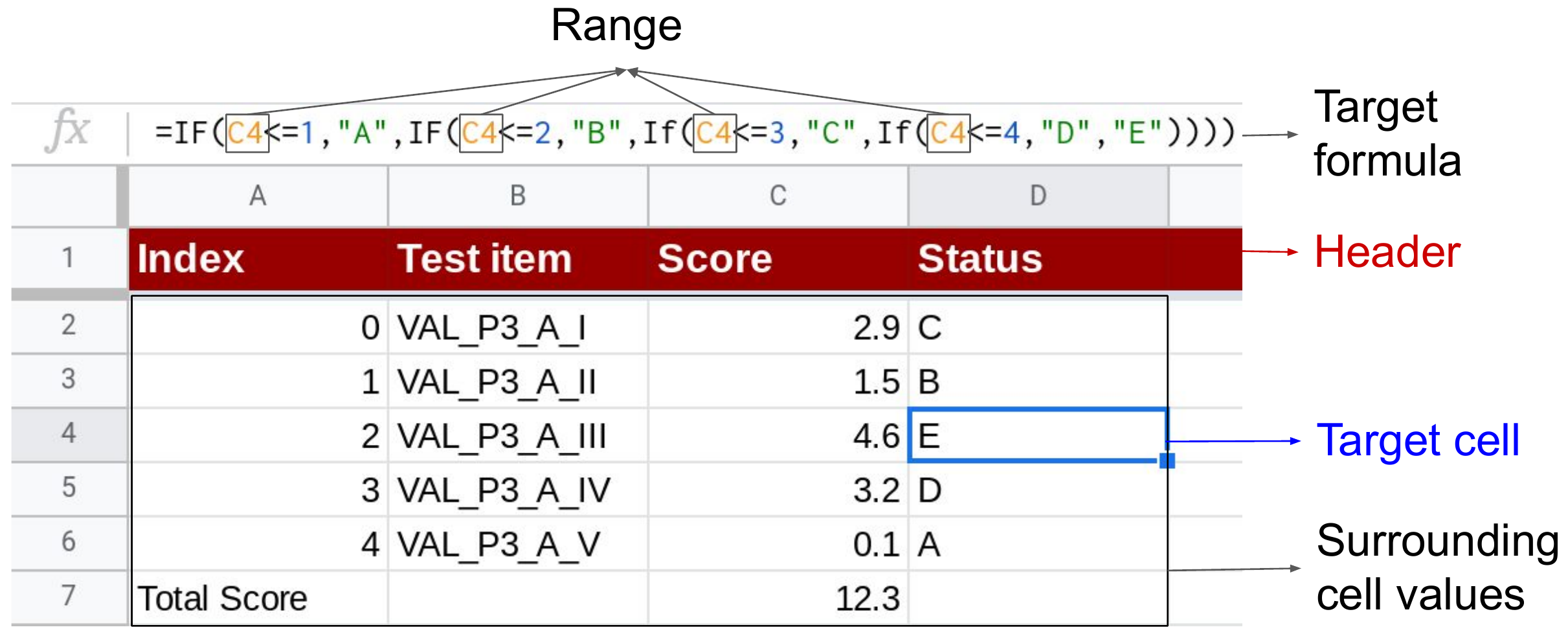}
    \caption{}
    \label{fig:sheets-ex-row}
    \end{subfigure}
    \begin{subfigure}[t]{0.45\linewidth}
    \includegraphics[width=\linewidth]{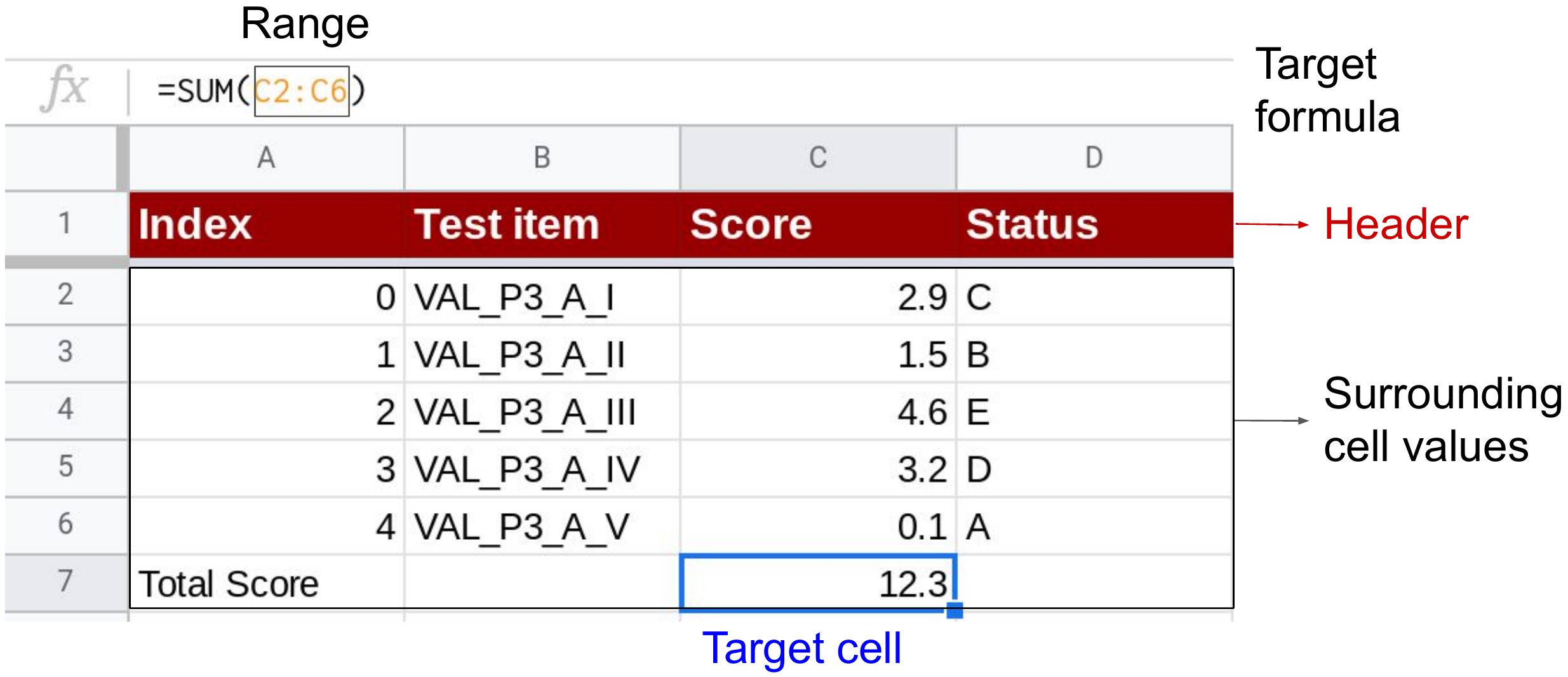}
    \caption{}
    \label{fig:sheets-ex-col}
    \end{subfigure}
    \begin{subfigure}[t]{0.45\linewidth}
    \includegraphics[width=\linewidth]{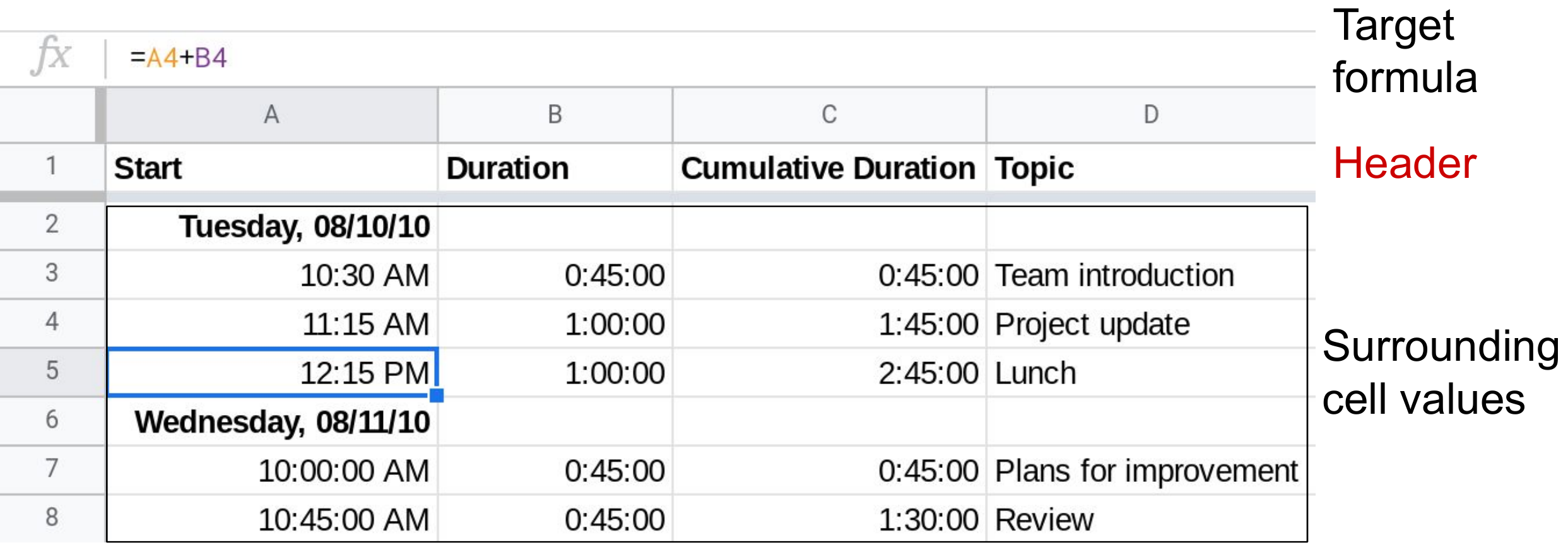}
    \caption{}
    \label{fig:sheets-ex-row-col-0}
    \end{subfigure}
    \begin{subfigure}[t]{0.45\linewidth}
    \includegraphics[width=\linewidth]{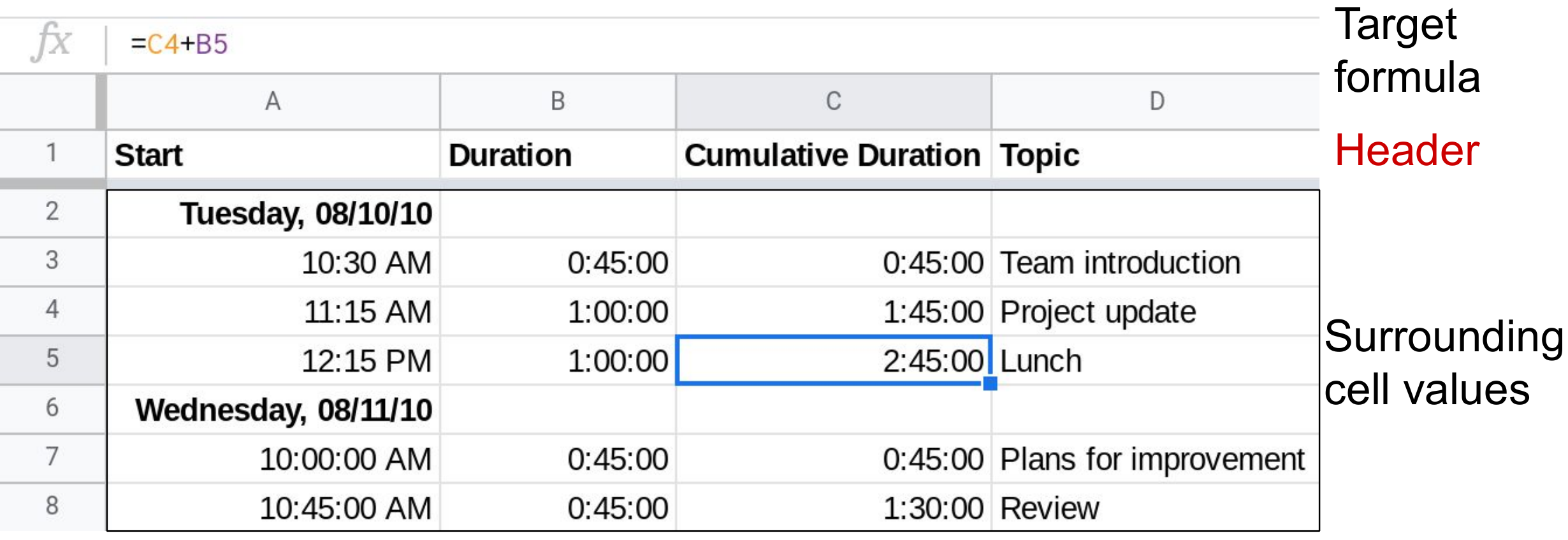}
    \caption{}
    \label{fig:sheets-ex-row-col-1}
    \end{subfigure}
    \caption{Illustrative synthetic examples of our spreadsheet formula prediction setup. (a): The formula manipulates cell values in the same row. (b): The formula is executed on the rows above. (c) and (d): Formulas involve cells in different rows and columns. The data value in the target cell is excluded from the input. All of these formulas can be correctly predicted by our model.}
    \label{fig:sheets-ex}
    \vspace{-0.5em}
\end{figure*}

\textbf{Input specification.} We illustrate the input context in Figure~\ref{fig:sheets-ex}.
The input context consists of two parts: (a) context surrounding the target cell (e.g., all cell values in rows 2--7, and columns A--D, excluding cell D4 in Figure~\ref{fig:sheets-ex-row}), and (b) the header row (e.g., row 1).
\charles{Add forward link to how we determine what the header is (I can do this once I read far enough to find it).}~\xinyun{It is in Appendix~\ref{app:implementation-details}. Is it an important detail that I should talk about in the main text? The header detection utilizes the Sheets format, i.e., the first row is considered as the header when it is fixed.}

In contrast to prior programming-by-example approaches~\citep{gulwani2011automating,parisotto2016neuro,devlin2017robustfill,vijayakumar2018neural}, our input specification features (a) tabular input, rather than independent rows as input-output examples, and (b) header information. Tabular input is important for many cases where formulas are executed on various input cells from different rows and columns (Figure~\ref{fig:sheets-ex}), and headers hold clues about the purpose of a column as well as its intended type, e.g, the header cell "Score" in Figure~\ref{fig:sheets-ex-col} is likely to indicate that the column data should be numbers.

\maniatis{Double check paragraph.} Note that we do not include the intended \emph{output} of the target cell in our input specification, for three reasons. First, unlike programming-by-example problems, we do not have multiple independent input-output examples available from which to induce a formula, so providing \emph{multiple} input-output examples is not an option. Second, even for our single input instance, the evaluated formula value may not be known by the spreadsheet user yet. Finally, we tried including the intended formula execution \emph{result} in our specification, but it did not improve the prediction accuracy beyond what the contextual information alone allowed.

\textbf{The spreadsheet language.} Our model predicts formulas written in the Google Sheets language\footnote{Google Sheets function list:~\url{https://support.google.com/docs/table/25273?hl=en}.}. Compared to the domain-specific language defined in FlashFill, which focuses on string transformations, the spreadsheet language supports a richer set of operators. Besides string manipulation operators such as~\texttt{CONCATENATE},~\texttt{LOWER}, etc., the spreadsheet language also includes operators for numerical calculations (e.g.,~\texttt{SUM} and~\texttt{AVERAGE}), table lookups (e.g., ~\texttt{VLOOKUP}) and conditional statements (\texttt{IF}, \texttt{IFS}). As will be discussed in Section~\ref{sec:exp}, around a hundred different base formula functions appear in our dataset, many more than the operators defined in the FlashFill DSL.

In this work, we limit our problem to formulas with references to \emph{local} cells in a spreadsheet tab, thus we exclude formulas with references to other tabs or spreadsheets, and absolute cell ranges. As will be discussed in Section~\ref{sec:approach}, we also exclude formulas with relative cell references outside a bounded range, i.e., farther than $D=10$ rows and columns in our evaluation. We consider improving the computational efficiency to support larger $D$ and enabling the synthesis of formulas with more types of cell references as future work.

\textbf{Formula representation.}
One of the key challenges in formula representation is how to represent cell references, especially ranges, which are prevalent in spreadsheet formulas. Naively using the absolute cell positions, e.g., \texttt{A5}, may not be meaningful across different spreadsheets. Meanwhile, a single spreadsheet can have millions of cells, thus the set of possible ranges is very large.

To address this,  we design a representation for formula sketches
inspired by prior work on sketch learning for program synthesis~\citep{solar2008program,murali2018neural,dong2018coarse,nye2019learning}.
A formula sketch includes every token in the prefix representation of the parse tree of the spreadsheet formula, except for cell references. References,
which can be either a single cell or a range of cells, are replaced with a special
placeholder~\texttt{RANGE} token. For example, the sketch of the formula in Figure~\ref{fig:sheets-ex-row} is \texttt{IF <= RANGE 1 "A" IF <= RANGE 2 "B" IF <= RANGE 3 "C" IF <= RANGE 4 "D" "E" \$ENDSKETCH\$}, where~\texttt{\$ENDSKETCH\$} denotes the end of the sketch. Notice that the sketch includes literals, such as the constants \texttt{1} and  \texttt{"A"}.

To complete the formula representation, we design an intermediate representation for ranges, \emph{relative} to the target cell, as shown in Figure~\ref{fig:range-grammar}. For example, \texttt{B5} in Figure~\ref{fig:sheets-ex-row-col-0} is represented as \texttt{\$R\$ R[0] C[1] \$ENDR\$} since it is on the next column but the same row as the target cell \texttt{A5}, and range \texttt{C2:C6} in Figure~\ref{fig:sheets-ex-col} is represented as \texttt{\$R\$ R[-5] C[0] \$SEP\$ R[-1] C[0] \$ENDR\$}.
The special tokens \texttt{\$R\$} and \texttt{\$ENDR\$} start and conclude a concrete range, respectively, and \texttt{\$SEP\$} separates the beginning and end (relative) references of a rectangular multi-cell range.

\begin{figure}[t]
\footnotesize
$$
\begin{array}{rcl}
\texttt{<Range>} & ::= & \texttt{\$R\$}~\texttt{<R>}~\texttt{<C>}~\texttt{\$ENDR\$} \\ 
& | & \texttt{\$R\$}~\texttt{<R>}~\texttt{<C>}~\texttt{\$SEP\$}~\texttt{<R>}~\texttt{<C>}~\texttt{\$ENDR\$}  \\
\texttt{<R>} & ::= & \texttt{R[-10]} \mid \texttt{R[-9]} \mid \texttt{...} \texttt{R[9]} \mid \texttt{R[10]} \\
\texttt{<C>} & ::= & \texttt{C[-10]} \mid \texttt{C[-9]} \mid \texttt{...} \texttt{C[9]} \mid \texttt{C[10]}
\end{array}
$$
\caption{The full grammar for range representation.}
\label{fig:range-grammar}
\vspace{-1.5em}
\end{figure}

A complete spreadsheet formula includes both the sketch and any concrete ranges; e.g., the formula in Figure~\ref{fig:sheets-ex-col} is represented as \texttt{SUM RANGE \$ENDSKETCH\$ \$R\$ R[-5] C[0] \$SEP\$ R[-1] C[0] \$ENDR\$ EOF}, where~\texttt{EOF} denotes the end of the formula. In Section~\ref{sec:decoder}, we will discuss our two-stage decoding process, which sequentially predicts the formula sketch and ranges.
\vspace{-0.8em}
\section{{\ours} Model Architecture}
\label{sec:approach}

\begin{figure}[t]
    \centering
    \includegraphics[width=\linewidth]{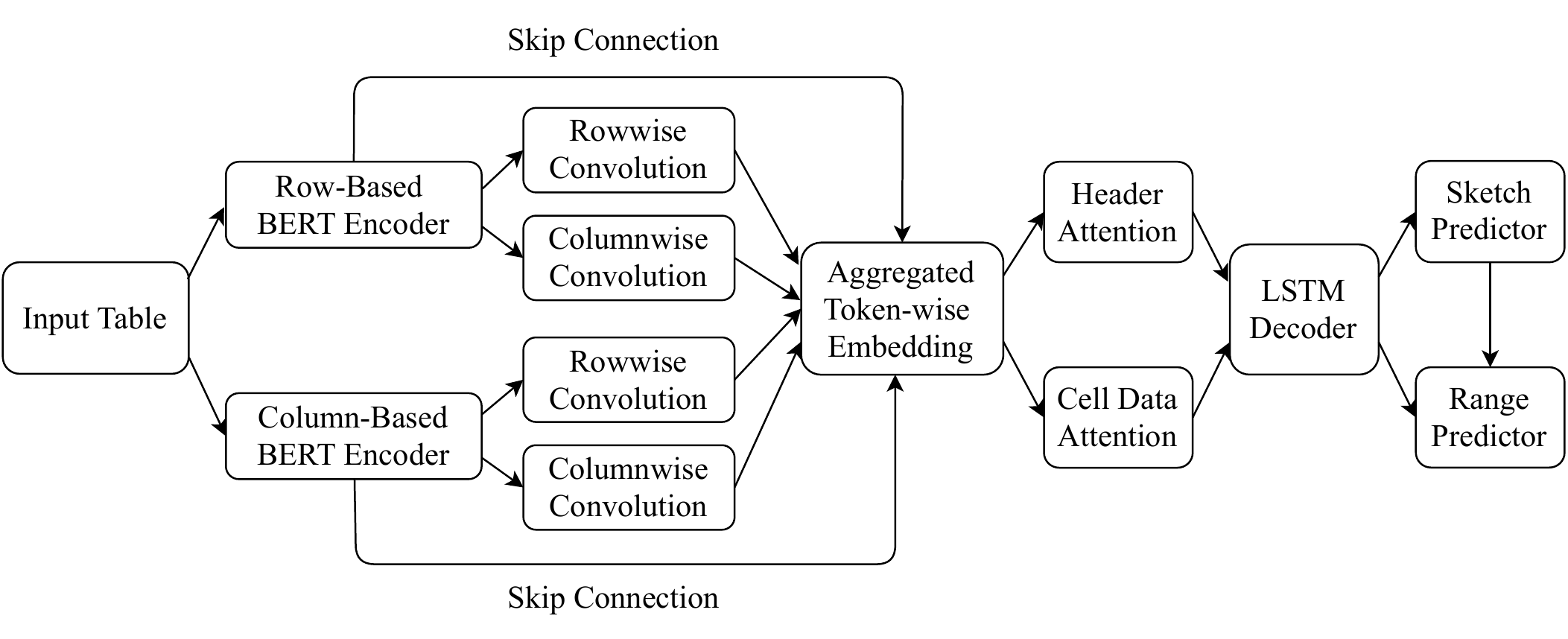}
    \caption{An overview of our model architecture.}
    \label{fig:model}
    \vspace{-2em}
\end{figure}

In this section, we present our{~\ours} model architecture for spreadsheet formula prediction. We provide an overview of our model design in Figure~\ref{fig:model}.

\subsection{Tabular Context Encoder}
\label{sec:encoder}

\textbf{Input representation.} Our model input includes the surrounding data values of the target cell as a table, and the first row is the header. When there is no header in the spreadsheet table, we set the header row to be an empty sequence. We include data values in cells that are at most $D$ rows and $D$ columns away from the target cell, so that the input dimension is $(2D+2) \times (2D+1)$, and we set $D=10$ in our experiments.

\textbf{Row-based BERT encoder.} We first use a BERT encoder~\citep{devlin2019bert} to compute a row-based contextual embedding for each token in the target cell's context.
Since our $2D+1+1$ rows contain many tokens and we use a standard BERT encoder of 512-token inputs, we \emph{tile} our rows into bundles of $N=3$ adjacent data rows, plus
the header row, which is included in every bundle.
Then we compute a token-wise BERT embedding for each bundle separately; the BERT weights are initialized from a pre-trained checkpoint for English. 
Specifically, in our experiments where $D=10$,
we concatenate all cell values for each row $i$ in the context into a token sequence
$R_i$, which has length $L=128$ (we trim and pad as needed). We combine rows in bundles $S_{rb}=[H_r, R_{3b-1}, R_{3b}, R_{3b+1}]$, for $b \in [-3, 3]$; here $H_r$ is the header row. We set the BERT segment IDs to 0 for the header tokens, and 1 for data tokens in each bundle. There are
$2D + 1 = 21$ rows of context, so each of the 21 data rows is
covered exactly once by the seven bundles.
The header row is assigned a different BERT representation 
in each bundle. To obtain a single representation of
the header row, we  average per token across the embeddings from all
of the bundles.

The number of data rows $N=3$ is set to seek the balance between the size of the tabular context fed into the encoder and the computational efficiency. Since the BERT we use takes 512 input tokens, we can feed at most $L=512/(N+1)$ tokens per row. To generate formulas referring to cells within $D=10$ rows and columns, $L=128$ is a good fit in our evaluation. If we further decrease $N$ and increase $L$, it imposes extra computational overhead due to more forward passes over BERT ($21/N$).

\textbf{Column-based BERT encoder.} As shown in Figure~\ref{fig:sheets-ex-col}, some formulas manipulate cells in the same column, in which case a column-based representation may be more desirable. Therefore, we also compute a column-based contextual embedding for all context tokens. We perform similar tiling as for the row-based BERT encoding, yielding
column bundles $S_{cb}$ for $b \in [-3, 3]$. Unlike with row-wise tiling, where we include the header row $H_r$ with every bundle, for column-wise tiling we use the column of the target cell, $H_c = C_0$, as the ``header column'' in every bundle.
After obtaining all token embeddings from this tiled computation by the BERT encoder, we discard token embeddings of $C_0$ in its role as header column, and only use its regular token embeddings from the bundle $S_{c0}$.

\textbf{Row-wise and column-wise convolution layers.} Although the output vectors of BERT encoders already contain important contextual information, such as headers, nearby rows and columns, they still do not fully embed the entire input table as the context. To encode the context from more distant rows and columns, we add a row-wise convolution layer and a column-wise convolution layer on top of each BERT encoder. Specifically, the row-wise convolution layer has a kernel size of $1\times L$, and the column-wise convolution layer has a kernel size of $(2D+2)\times 1$ for row-based BERT, and $(2D+1)\times 1$ for column-based BERT. In this way, the convolution layer aggregates across BERT embeddings from different bundles,
allowing the model to take longer range dependencies into account.\charles{Xinyun: The previous sentence is new. Does this sound good to you?}~\xinyun{I moved this sentence here, thanks!} For each input token, let $e_b$ be its BERT output vector, $c_r$ be the output of the row-wise convolution layer, and $c_c$ be the output of the column-wise convolution layer. 
The final embedding of each input token is the concatenation of the BERT output and the output of convolution layers, i.e.,  $e=[c_r+c_c; e_b]$.

\vspace{-0.5em}
\subsection{Two-stage Formula Decoder}
\label{sec:decoder}

We train an LSTM~\citep{hochreiter1997long} decoder to generate the formula as a token sequence. Meanwhile, we use the standard attention mechanism~\citep{bahdanau2014neural} to compute two attention vectors, one over the input header, and one over the cell data. We concatenate these two attention vectors with the LSTM output, and feed them to a fully-connected layer with the output dimension $|V|$, where $|V|$ is the vocabulary size of formula tokens. Note that the token vocabularies are different for sketches (formula operators, literals, and special tokens) and ranges (relative row and column tokens and special range tokens). The output token prediction is computed with the softmax.

As mentioned in Section~\ref{sec:setup}, we design a two-stage decoding process, where the decoder first generates the formula sketch, and then predicts the concrete ranges. In the first stage, the 
sketch is predicted as a sequence of tokens by the LSTM,
and the prediction terminates when an~\texttt{\$ENDSKETCH\$} token is generated.
Then in the second stage, the range predictor sequentially generates formula ranges corresponding to each~\texttt{RANGE} token in the sketch, and the prediction terminates when an~\texttt{EOF} token is generated. Both sketch and range predictors share the same LSTM, but with different output layers.
\charles{I was confused about this sentence: "the decoder uses another fully-connected layer to generate ranges. Specifically,". Is it correct to say: "The ranges are also predicted as a sequence of tokens by an LSTM with the same weights the sketch predictor, except for the final
output layer, which is different because of the different
vocabularies between sketches and ranges."}~\xinyun{It is technically correct. I revised the sentence, is it clearer?}~\charles{Yes, this is very clear now. Thank you!}
\section{Experiments}
\label{sec:exp}

We evaluate {\ours} on spreadsheet formula prediction tasks in different settings. We first describe our dataset, then introduce our experimental setup and discuss the results~\footnote{The code and data are available at \url{https://github.com/google-research/google-research/tree/master/spreadsheet_coder}.}.

\subsection{Dataset}

We constructed our dataset from a corpus of Google Sheets publicly shared within our organization. We collected 46K Google Sheets with formulas, and split them into 42K for training, 2.3K for validation, and 1.7K for testing.

Although in principle, our model could generate formulas using any operator in the spreadsheet language, some kinds of value references are impossible to predict from local context, thus we remove formulas with such values from our dataset. Specifically, we exclude formulas that use the \texttt{HYPERLINK} function with a literal URL, since those are merely "stylistic" formulas that perform no computation beyond presenting a URL as a clickable link. As discussed in Section~\ref{sec:setup}, we also filtered out formulas with cross-references from other tabs or spreadsheets, with cell references farther than 10 rows or columns from the target cell in either direction, or with absolute cell ranges. Finally, our dataset includes 770K training samples, 42K for validation, and 34K for testing.

About the length distribution of target spreadsheet formulas, about $32\%$ formulas have sketch lengths of 2, $53\%$ formulas have sketch lengths of 3, $11\%$ formulas have sketch lengths of 4-5, and $4\%$ formulas have sketch lengths of at least 6. As discussed in Section~\ref{sec:setup}, even if the formula sketches are mostly short, it is still challenging to generate the full formulas correctly. For example, the formula in Figure~\ref{fig:sheets-ex-col} is represented as \texttt{SUM RANGE \$ENDSKETCH\$ \$R\$ R[-5] C[0] \$SEP\$ R[-1] C[0] \$ENDR\$ EOF}, which has a sketch length of 2, but the full formula length is 10 if excluding the~\texttt{EOF} token for length calculation. In total, around a hundred operators are covered in our output vocabulary, including 82 spreadsheet-specific functions, and other general-purpose numerical operators (e.g.,~\texttt{+},~\texttt{-}). We defer more details about dataset construction process and dataset statistics to Appendix~\ref{app:data}.

By default, each sample includes both the header row and surrounding data values of relative cell positions within $[-10, 10]$. Note that we do not include the data of the target cell, and we leave an empty value there. We perform the header detection according to the spreadsheet table format, i.e., we recognize the first row of a table as the header when it is frozen. Though some spreadsheet tables may include header-like descriptions in the leftmost column, e.g., ``Total Score'' in Figure~\ref{fig:sheets-ex-row}, we only extract headers as a row, to ensure the precision of header detection. In Section~\ref{sec:exp-results}, we also discuss settings when the model input does not include headers, and when we only include a few data rows above the target cell as the input context.

\vspace{-1em}
\subsection{Evaluation Setup}

\textbf{Metrics.} We evaluate the following metrics: (1) \emph{Formula accuracy}: the percentage of predicted formulas that are the same as the ground truth. (2) \emph{Sketch accuracy}: the percentage of predictions with the same formula sketches as the ground truth. As discussed in Section~\ref{sec:setup}, formula sketches do not include ranges, but include both functions and literals. (3) \emph{Range accuracy}: the percentage of predictions with the same ranges as the ground truth. Note that the order of predicted ranges should also be the same as the ground truth. In addition, the model may predict the ranges correctly even if the sketch prediction is wrong, as shown in Figure~\ref{fig:sheets-wrong-sketch-ex}.\eat{ \maniatis{Is this conditioned on the sketch being correct? In other words, is formula accuracy == sketch accuracy * range accuracy?}~\xinyun{No, these two metrics are independently computed. See Figure~\ref{fig:sheets-wrong-sketch-ex} for an example of the correct range with the wrong sketch.} \charles{Xinyun, based
on your answer to Petros, do you think it would make sense 
for us to add: "It is possible for the model to predict
the ranges correctly even if the sketch is predicted incorrectly,
for example, see Figure~\ref{fig:sheets-wrong-sketch-ex}. 
}~\xinyun{Added this explanation. I didn't refer to the figure earlier because I already put it in the appendix.}}

Note that our formula accuracy metric could be an underestimate of the semantic equivalence, because different spreadsheet formulas may be semantically equivalent. For example, to predict arguments for~\texttt{SUM} and~\texttt{MULTIPLY}, different orders of the cell ranges have the same meaning. However, it is hard to systematically define the semantic equivalence in our evaluation, because we aim to support a wide range of operators in the spreadsheet language. Some existing works on program synthesis have evaluated the semantic equivalence based on the execution results~\citep{devlin2017robustfill,bunel2018leveraging,sun2018neural}. However, it is hard to sample different input spreadsheets requiring the same formula, thus evaluating the execution accuracy is challenging. Therefore, we still focus on our current metric to measure the formula accuracy, where we compare whether the predicted formula is exactly the same as the single ground truth formula included in the spreadsheet.

\textbf{Model details.} For models with the BERT encoder~\citep{devlin2019bert}, including our full{~\ours} model, we use the BERT-Medium architecture, and initialize from the 
English pre-trained model by default.\footnote{We downloaded the pre-trained BERT from:~\url{https://github.com/google-research/bert}.} \eat{\charles{Xinyun: Can you confirm that it is the English
checkpoint that we used?}~\xinyun{Yes.}} We compared our full model with several variants:

{(1) Different encoder architectures}.  i) Using a single BERT encoder, either row-based or column-based; ii) removing convolution layers, where the BERT output is directly fed into the decoder.

{(2) Different decoding approaches}. We compare our two-stage decoding discussed in Section~\ref{sec:decoder}
to a simpler model that uses the same predictor for both the sketch and ranges, with a single joint output vocabulary for both.\eat{\maniatis{The figure isn't showing the fully-connected layer. Do you mean that there was no autoregressive decoder? Or that there was a single autoregressive decoder, and ranges were predicted directly within the formula? }~\xinyun{Modified the description a bit. Basically, it means that we have a single predictor for both sketch and ranges, instead of two separate predictors.}} \charles{I am not sure what two predictors vs one means. Does one predictor mean
that there are two different LSTMs instead of one?}~\xinyun{Your understanding in Section~\ref{sec:decoder} it correct. In both cases, they use the same LSTMs, and the only different comes from the output layers.}
\rishabh{should we also mention the variant here where we do not perform two-level decoding but predict the formula with ranges all at once?}~\xinyun{I tried it a bit, and the performance is similar to this setting.}

{(3) Different model initialization}.  When not using the pre-trained BERT model weights, we randomly initialize BERT encoders. This tests whether pre-training on generic natural
language text is useful for our spreadsheet data.

We compare to previous approaches for related program synthesis tasks. First, we evaluate RobustFill, which demonstrates the state-of-the-art performance on string manipulation tasks for Excel spreadsheets~\citep{devlin2017robustfill}. Specifically, RobustFill encodes the cell context as independent rows, rather than a 2D table as in SpreadsheetCoder. Afterwards, at each decoding step, a shared LSTM decoder generates a hidden state per data row, which are then fed into a max pooling layer. Finally, the pooled hidden state is fed into a fully-connected layer to predict the formula token. We trained two variants of RobustFill on our dataset: one encodes each row independently, and another encodes each column independently, denoted as~\emph{row-based RobustFill} and~\emph{column-based RobustFill} respectively. In addition, we compared to a baseline that does not utilize any input context, thus the model only includes the LSTM decoder, similar to prior work on language modeling~\citep{sundermeyer2012lstm,karpathy2015visualizing}.

\vspace{-0.5em}
\subsection{Results}
\label{sec:exp-results}
In this section, we present the results using different variants of spreadsheet contexts as the model inputs. We perform a beam search during the inference time. Empirically, we find that results with different beam sizes (2, 4, 8, 16, 32, 64, 128) are similar, i.e., the accuracies vary within $0.3\%$. Therefore, we set the beam size to be 64 for all settings.

\vspace{-0.5em}
\subsubsection{Results with the Full Input Context}
\label{sec:exp-results-full-context}
\vspace{-0.5em}
Using both headers and the full surrounding data cell values as the model input, we present the formula accuracy in Table~\ref{tab:res-full}, where top-$k$ accuracy measures 
how often the ground truth appears in the top $k$ predictions using  beam search.
Compared to the model without the input context, all other models are able to use the contextual data to provide more accurate predictions. In particular, our full model achieves over $40\%$ top-1 full formula prediction accuracy, which is 4 times as high as the model without context. We also observe
that the full SpreadsheetCoder model has
much better accuracy than either of the RobustFill models,
demonstrating that our model is more capable of leveraging the implicit specification provided by the tabular context.

\textbf{Different encoder architectures.} Appropriately encoding the input context is important. Comparing with RobustFill models, we observe that it is beneficial to model the dependency among different rows and columns, instead of encoding each row or column independently. Meanwhile, adding convolution layers brings additional performance gain, because it enables the representation of each input token to aggregate broader contextual information beyond a few nearby rows or columns, i.e., 3 for our BERT encoders as discussed in Section~\ref{sec:encoder}.\charles{I don't believe the explanation that we gave in the previous sentence, because the BERT models already allow the representation of each input token to rely on broader contextual information. Is there any way we can improve this explanation? (It's not obvious to me.) I'm also OK if we remove the phrase "because it enables...."}\xinyun{Not really. With the BERT model, it only aggregates the context from the nearby 3 rows or columns, as discussed in Section~\ref{sec:encoder}. By adding CNN, it allows the embedding to aggregate information from all other rows or columns. I added a short explanation, is it clearer?}\charles{OHHHH! I UNDERSTAND! It's because the CNN aggregates across bundles. Let's leave this, then. I have added a phrase to emphasize this in Section 3. I've highlighted it for you with a comment.}~\xinyun{That sentence looks good, thanks!}
Finally, although models representing the input context as column-based tables generally perform worse than those using row-based tables, including both row-based and column-based encoders improves the overall accuracies by $2$--$3$ percentage points. Note that the improvement is not due to the larger model size: to test this, we trained row-based and column-based BERT models with the larger BERT-base and BERT-large architectures, but the results were no better, while taking longer to train. In addition, initializing from pre-trained BERT encoders increases the formula accuracy by around $10$ percentage points, suggesting that although spreadsheet headers are generally short natural language phrases, pre-training on a large-scale text corpus with much more complex text still enables the model to better understand the spreadsheet context. \maniatis{Can we also say this? "Furthermore, natural language also appears in non-header data cells, motivating the use of BERT encoding for all cells."}\xinyun{Earlier I would like to put in this way, but since pre-training doesn't help in the no-header setting, I am not sure if we want to emphasize this point.}

\begin{figure}[t]
    \centering
    \begin{subfigure}[t]{\linewidth}
    \includegraphics[width=\linewidth]{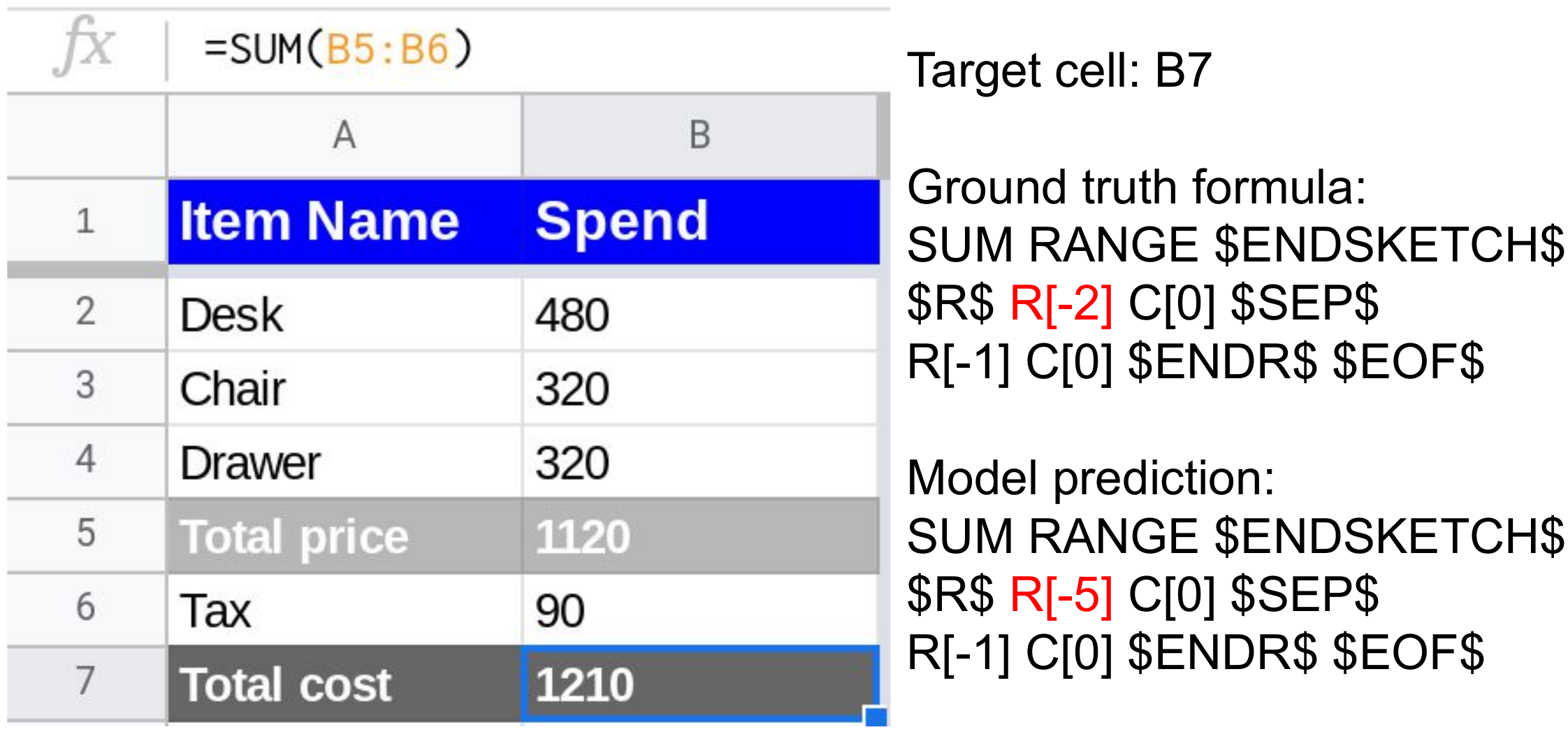}
    \caption{}
    \label{fig:sheets-wrong-range-ex}
    \end{subfigure}
    \begin{subfigure}[t]{\linewidth}
    \includegraphics[width=\linewidth]{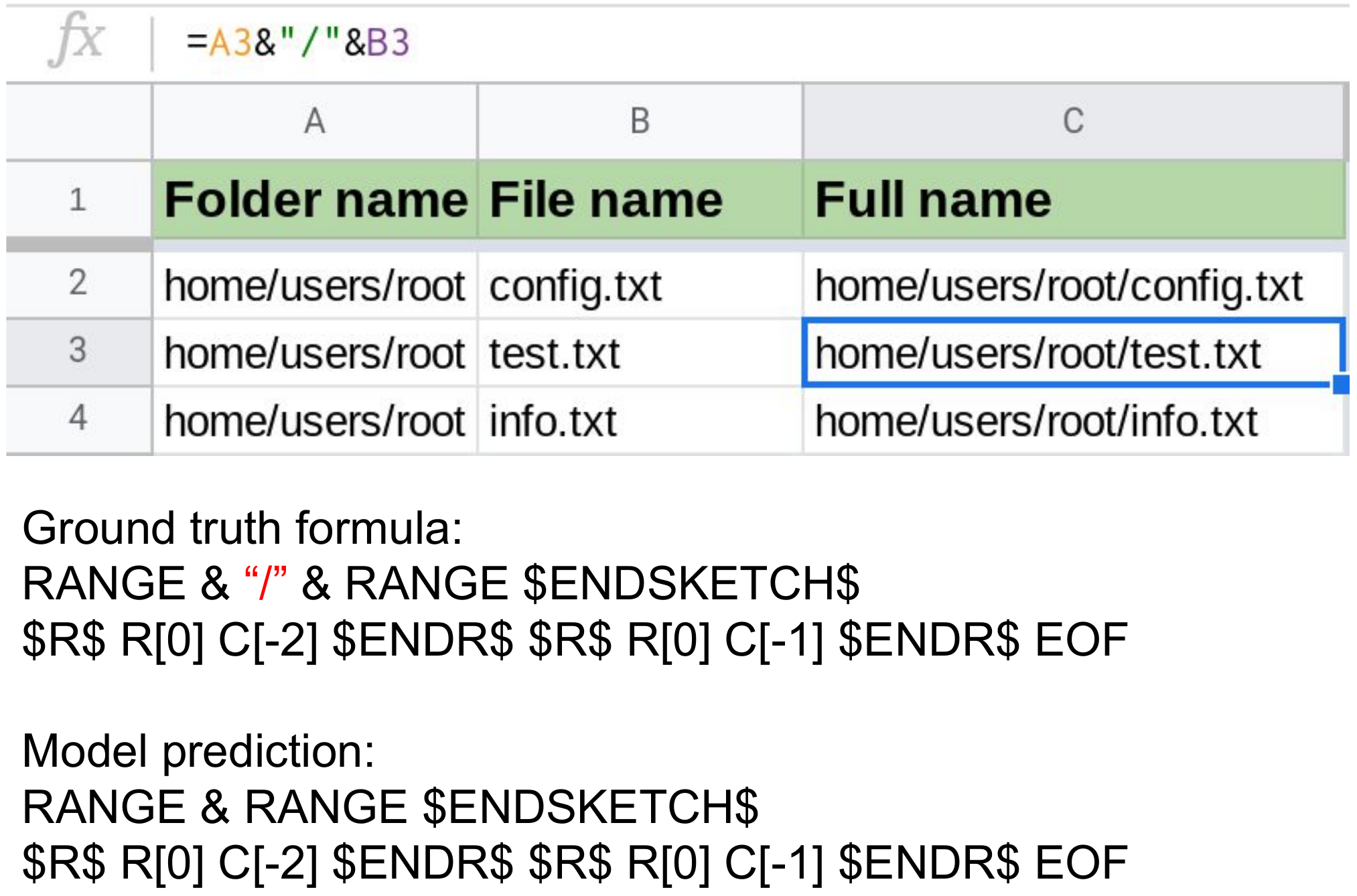}
    \caption{}
    \label{fig:sheets-wrong-sketch-ex}
    \end{subfigure}
    \caption{Examples of wrong formula predictions by our full model. (a) The sketch prediction is correct, but the range is wrong. (b) The range prediction is correct, but the sketch is wrong. These are synthetic examples for illustrative purposes.}
    \label{fig:sheets-wrong-ex}
    \vspace{-2em}
\end{figure}

\textbf{Breakdown analysis of sketch and range prediction.} We present the sketch and range accuracies in Table~\ref{tab:res-breakdown}. On the one hand, sketch accuracies are generally much higher than range accuracies, since formulas are more likely to share common sketches with similar spreadsheet context, while range prediction requires a more careful investigation of the table structure. On the other hand, sketch prediction becomes more challenging when literals are included. In Figure~\ref{fig:sheets-wrong-range-ex}, we present a prediction with the correct sketch but the wrong range. Specifically, the model could easily infer that the formula should call a~\texttt{SUM} function, since it is a common prediction given the input token ``Total''. However, the model wrongly selects all cells above as the function argument, and ignores the fact that the cell \texttt{B5} is already the sum of cells \texttt{B2}--\texttt{B4}, indicated by the text ``Total price'' in cell \texttt{A5}. Figure~\ref{fig:sheets-wrong-sketch-ex} shows a prediction with the correct range but the wrong sketch, where the predicted formula misses a ``/'' as an argument to the string concatenation operator ``\&''.  Two-stage decoding disentangles the generation of sketches and ranges, so that the two predictors could focus on addressing different difficulties in formula prediction, and this mechanism improves the overall accuracy.

\textbf{Prediction on formulas with different sketch lengths.} We present the top-1 formula accuracy on formulas with different sketch lengths in Figure~\ref{fig:res-sketch-length}. Note that we exclude the~\texttt{\$ENDSKETCH\$} token from length calculation. First, all models achieve higher performance on formulas with sketch lengths of $2$--$3$ than longer formulas. It is harder to make exactly the same prediction as the ground truth when the formula becomes longer, especially given that the input context is often an ambiguous specification for formula prediction. Fortunately, users typically do not need to write complicated formulas for spreadsheet data manipulation. Specifically, 85\% of our collected formulas have sketch lengths of $2$--$3$. Despite the performance degradation, our full model consistently performs better than other models on formulas with different sketch lengths.

\begin{table}[t]
\begin{minipage}{\linewidth}
    \caption{Formula accuracy on the test set. ``$-$'' means the corresponding component is removed from our full model.}
    \label{tab:res-full}
    \centering
    \scalebox{0.85}{
\begin{tabular}{rlccc}
    \toprule
& \textbf{Approach} & \textbf{Top-1} & \textbf{Top-5} & \textbf{Top-10} \\    \midrule
       &Full Model & \textbf{42.51\%} & \textbf{54.41\%} & \textbf{58.57\%} \\
$-$    &Column-based BERT & 39.42\% & 51.68\% & 56.50\% \\
$-$    &Row-based BERT & 20.37\% & 40.87\% & 48.37\% \\
$-$    &Convolution layers & 38.43\% & 51.31\% & 55.87\% \\
$-$    &Two-stage decoding & 41.12\% & 53.57\% & 57.95\% \\
$-$    &Pretraining & 31.51\% & 42.64\% & 49.77\% \\    \midrule
    &Row-based RobustFill & 31.14\% & 40.09\% & 47.10\% \\
    &Column-based RobustFill & 20.65\% & 39.69\% & 46.96\% \\
    &No context & 10.56\% & 23.27\% & 31.96\% \\    \bottomrule
    \end{tabular}
    }
\end{minipage}
\begin{minipage}{\linewidth}
    \centering
    \includegraphics[width=1.1\linewidth]{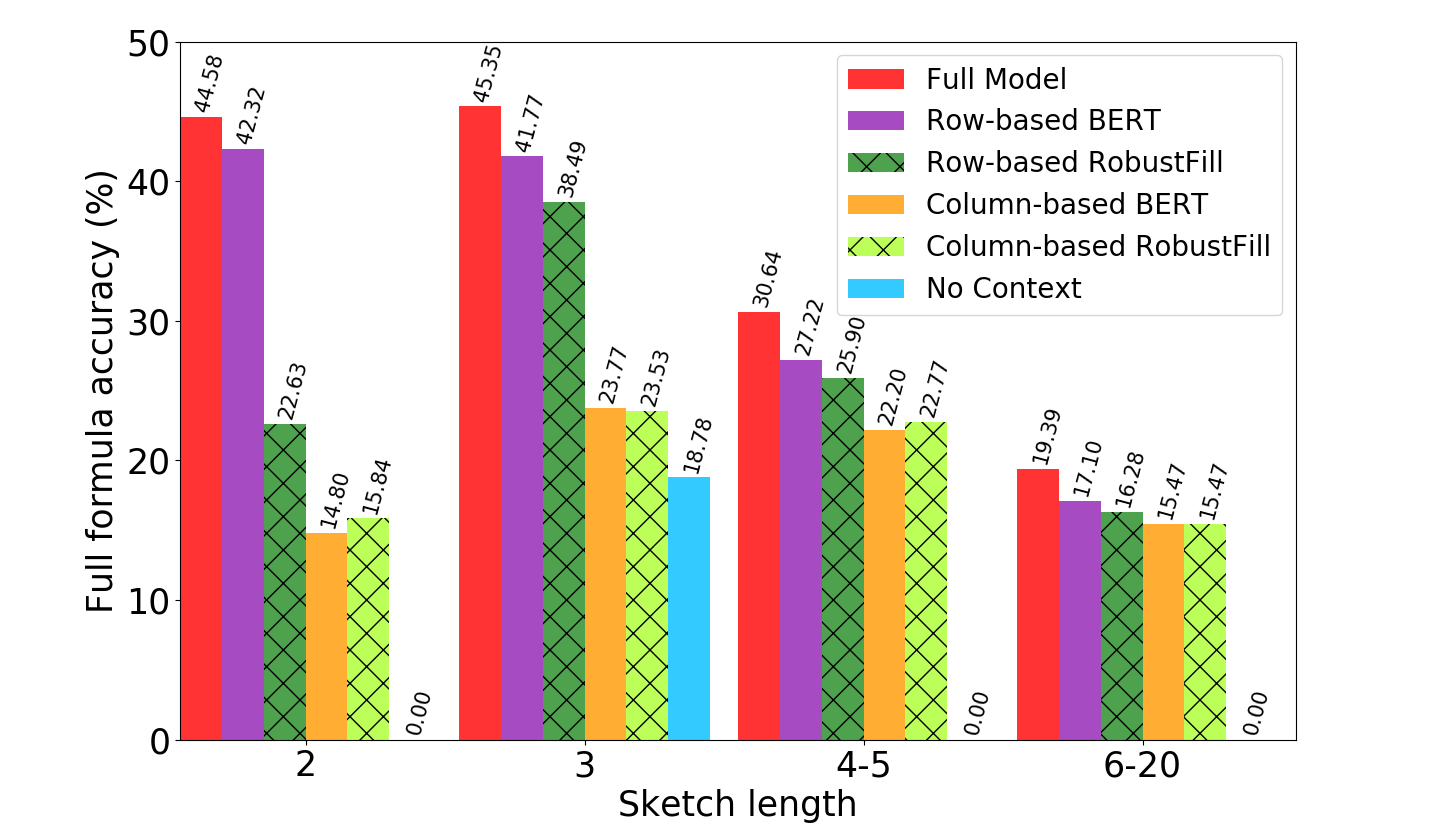}
    \captionof{figure}{Top-1 formula accuracies for different sketch lengths.}
    \label{fig:res-sketch-length}
\end{minipage}
\vspace{-2em}
\end{table}

\begin{table}[t]
\caption{Sketch and range accuracy on the test set.}
\label{tab:res-breakdown}
\begin{subtable}{\linewidth}
    \caption{Sketch accuracy.}
    \label{tab:res-sketch}
    \centering
    \scalebox{0.85}{
\begin{tabular}{rlccc}
    \toprule
    & \textbf{Approach}  & \textbf{Top-1} & \textbf{Top-5} & \textbf{Top-10} \\
    \midrule
    &Full Model & \textbf{57.41\%} & \textbf{72.04\%} & \textbf{78.52\%}  \\
    $-$ &Column-based BERT & 55.50\%  & 70.88\%  & 77.73\%  \\
    $-$ &Row-based BERT & 27.49\%  & 61.95\%  & 73.95\%  \\
    $-$ &Convolution layers & 53.68\%  & 69.38\% & 75.67\%  \\
    $-$ &Two-stage decoding & 56.47\%  & 72.02\%  & 78.30\%  \\
    $-$ &Pretraining & 41.26\% & 64.67\% & 76.36\% \\
    \midrule
    &Row-based RobustFill & 40.23\% & 61.50\%  & 72.20\% \\
    &Column-based RobustFill & 29.50\%  & 59.97\% & 71.31\% \\
    &No context & 25.19\% & 47.08\% & 52.70\% \\
    \bottomrule
    \end{tabular}}
\end{subtable}
\begin{subtable}{\linewidth}
    \caption{Range accuracy.}
    \label{tab:res-range}
    \centering
    \scalebox{0.85}{\begin{tabular}{rlccc}
    \toprule
    & \textbf{Approach}  & \textbf{Top-1} & \textbf{Top-5} & \textbf{Top-10} \\
    \midrule
    & Full Model & \textbf{46.93\%} & \textbf{59.60\%} & \textbf{63.51\%}  \\
    $-$ &Column-based BERT & 43.60\% & 57.12\%  & 62.27\% \\
    $-$ &Row-based BERT & 22.57\% & 47.84\% & 55.29\% \\
    $-$ &Convolution layers & 42.84\%  & 56.64\% & 61.03\%  \\
    $-$ &Two-stage decoding & 44.59\% & 58.52\%  & 62.48\% \\
    $-$ &Pretraining & 36.03\% & 49.85\% & 54.71\% \\
    \midrule
    &Row-based RobustFill & 33.88\% & 48.16\% & 54.83\% \\
    &Column-based RobustFill & 23.97\% & 47.09\% & 52.75\% \\
    &No context & 11.80\% & 25.54\% & 38.07\% \\
    \bottomrule
    \end{tabular}}
\end{subtable}
\vspace{-2em}
\end{table}

\vspace{-0.5em}
\subsubsection{The Effect of Header Information}
\label{sec:exp-results-wo-headers}

\begin{table}[t]
\begin{minipage}{\linewidth}
    \caption{Formula accuracy on the test set, excluding headers in the context. Corresponding results with headers are in Table~\ref{tab:res-full}.\maniatis{What is the line about RobustFill with headers? DO you give header tokens to RobustFill with every input output example? Is this described in the text?}~\xinyun{Results of RobustFill with headers are in Table~\ref{tab:res-full}.}~\charles{Xinyun: I added your comment to the caption. Does this seem clear to you?}~\xinyun{Yes it is clear, thanks!}}
    \label{tab:res-full-exclude-headers}
    \centering
    \scalebox{0.85}{
    \begin{tabular}{rlccc}
    \toprule
    & \textbf{Approach} & \textbf{Top-1} & \textbf{Top-5} & \textbf{Top-10} \\
    \midrule
    &Full Model & 20.47\% & 40.23\% & 47.40\% \\
    $-$ &Column-based BERT & 20.63\% & 40.40\% & \textbf{48.70\%} \\
    $-$ &Row-based BERT & 20.38\% & 40.11\% & 47.88\% \\
    $-$ &Pretraining & \textbf{20.94\%} & \textbf{40.64\%} & 48.51\% \\
    \midrule
    &Row-based RobustFill & 19.02\% & 33.60\% & 37.38\% \\
    &Column-based RobustFill & 17.64\% & 30.45\% & 36.79\% \\
    &No context & 10.56\% & 23.27\% & 31.96\% \\
    \bottomrule
    \end{tabular}}
\end{minipage}
\begin{minipage}{\linewidth}
    \centering
    \includegraphics[width=\linewidth]{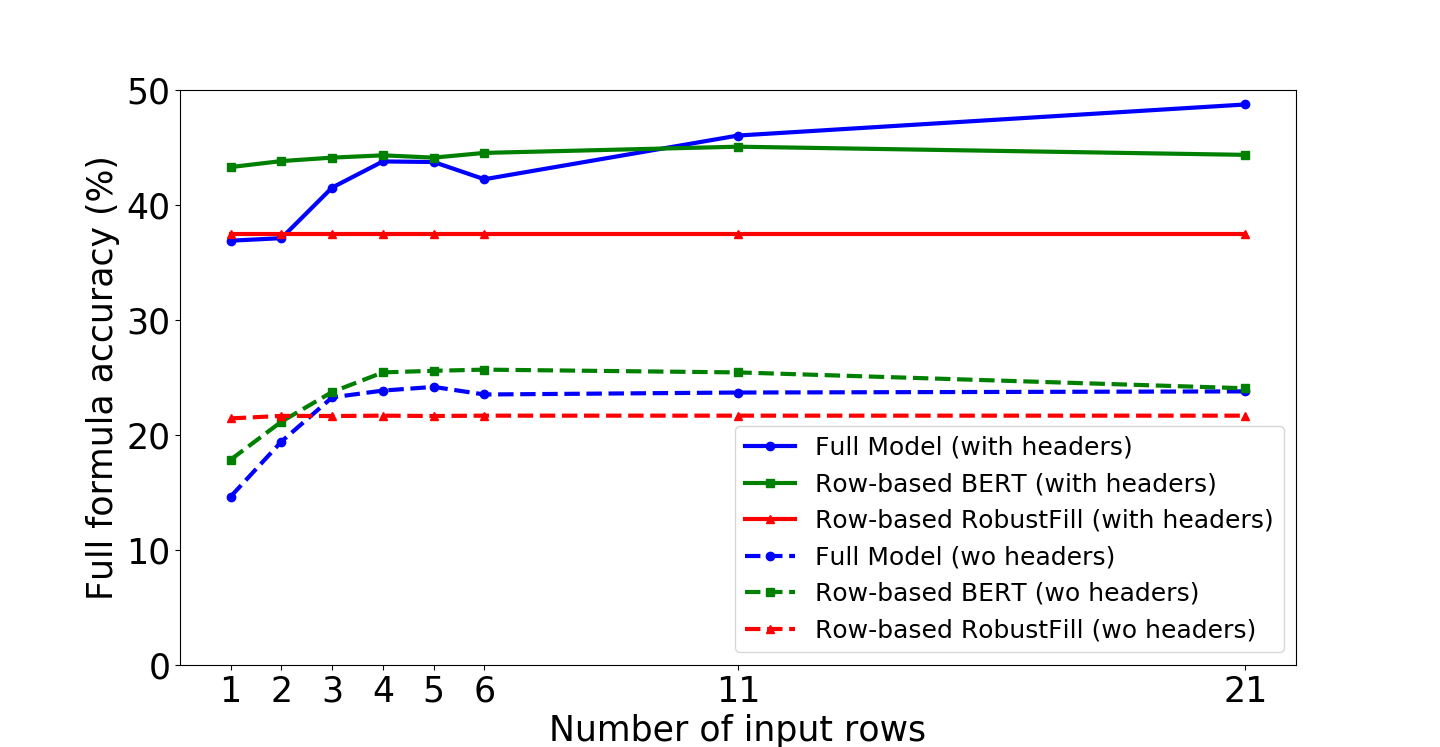}
    \captionof{figure}{Top-1 formula accuracy in the FlashFill-like setting, with different number of input rows.}
    \label{fig:res-flashfill}
\end{minipage}
\end{table}

\begin{figure}[t]
    \centering
    \begin{subfigure}[t]{\linewidth}
    \includegraphics[width=\linewidth]{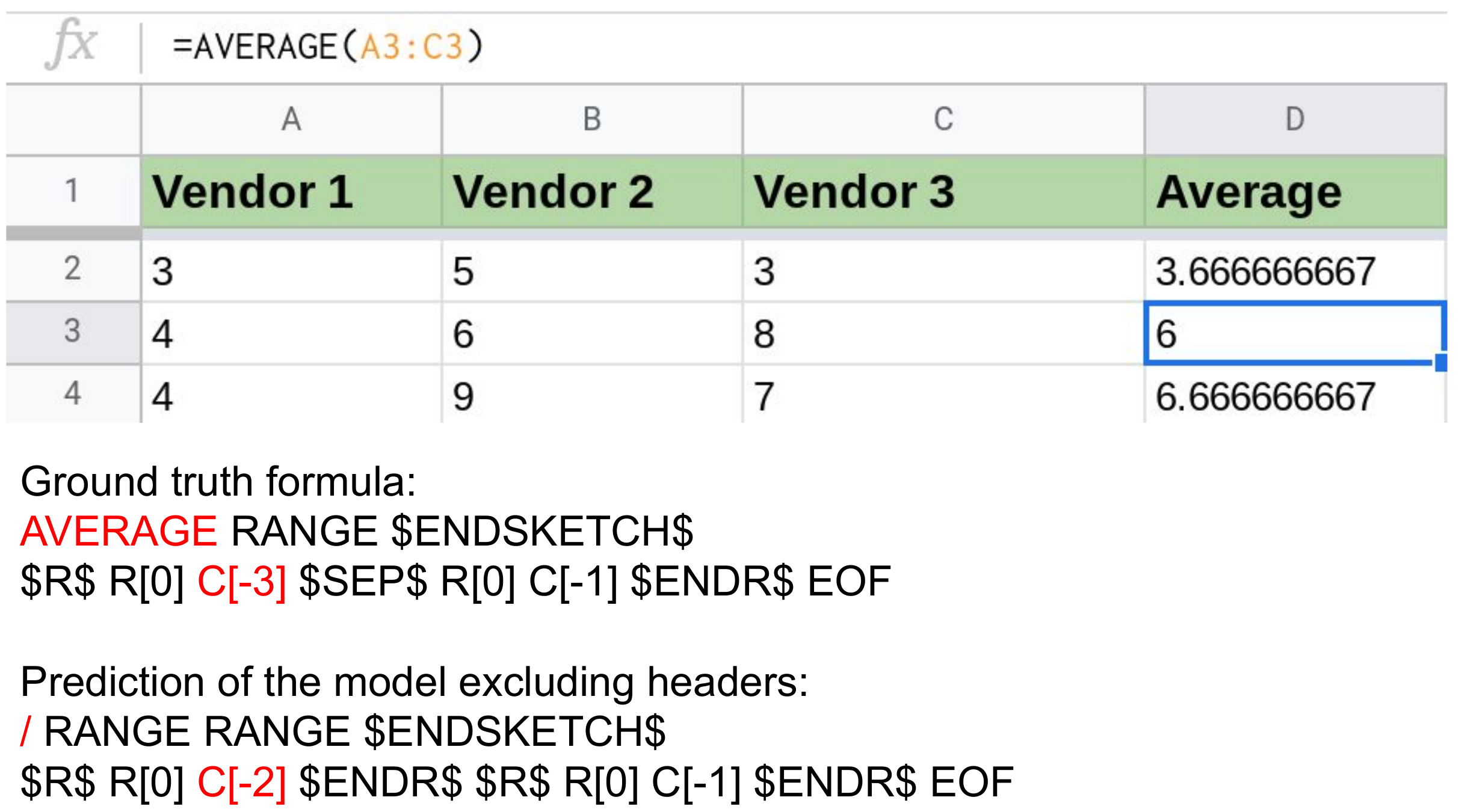}
    \caption{}
    \label{fig:sheets-wrong-wo-headers-ex}
    \end{subfigure}
    \begin{subfigure}[t]{\linewidth}
    \includegraphics[width=\linewidth]{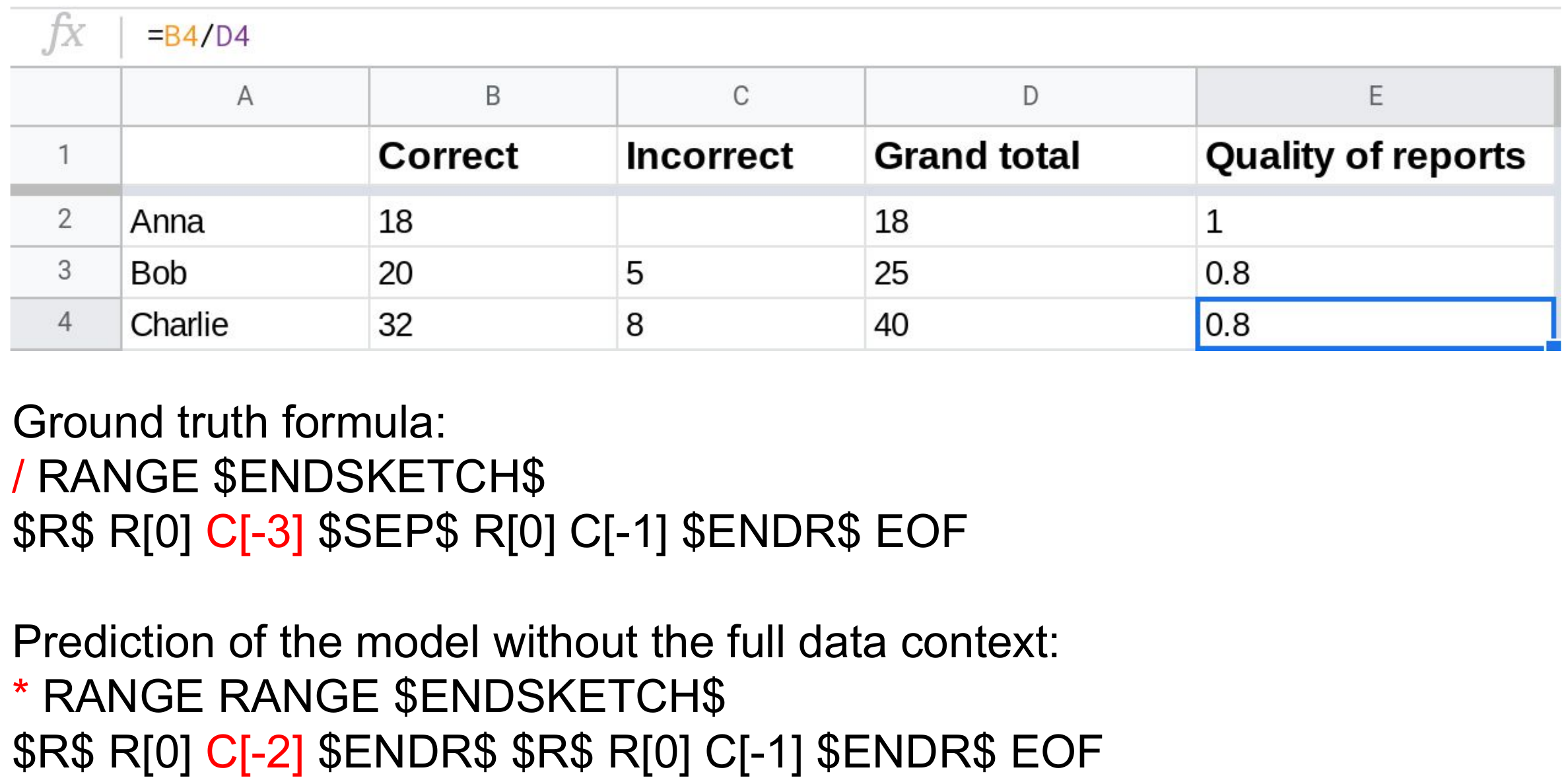}
    \caption{}
    \label{fig:sheets-wrong-wo-data-ex}
    \end{subfigure}
    \caption{Examples of formulas that are correctly predicted by our full model with the full context, but wrongly predicted with missing context. (a) The wrong prediction when the model input does not include headers. Note that the model with headers predicts it correctly even if only one data row is provided. (b) The wrong prediction when the model input only includes headers and one data row. These are synthetic examples for illustrative purposes.}
    \label{fig:sheets-wrong-missing-context-ex}
    \vspace{-1em}
\end{figure}
In this section, we evaluate the effect of including the header row as the model input, which usually provides a short description of the table in natural language. 
For all models, we remove the headers from the context
by replacing the header tokens with empty values. Thus the models can
only use surrounding data cells as the spreadsheet context.

In Table~\ref{tab:res-full-exclude-headers}, we observe a notable accuracy drop compared to Table~\ref{tab:res-full}, indicating that leveraging headers is critical. Figure~\ref{fig:sheets-wrong-wo-headers-ex} shows an example that can be correctly predicted by our full model, but is wrongly predicted by the model without input headers. We can observe that without the header ``Average'', it is much harder to figure out that the formula should call the~\texttt{AVERAGE} function instead of a division.
Interestingly, without input headers, using row-based or column-based table representation no longer makes much difference. However, our tabular input context encoders still perform better than RobustFill models, suggesting the importance of modeling the dependency among different rows and columns. In addition, initializing from pre-trained BERT model weights does not improve the results, and even slightly hurts the performance. The main reason is that the cell data values are mostly numeric and string literals. Breakdown results are deferred to Appendix~\ref{app:results}.

\subsubsection{Results in the FlashFill-like Setting}

In this section, we conduct experiments in the FlashFill-like setting, where formulas are always executed on cells in the same row. In total, 2.5K formulas in the test set only include cells with the relative row position~\texttt{R[0]}, which constitute around $73\%$ of the test set. More details are in Appendix~\ref{app:flashfill}.
\charles{Can we expand on what "formulas of this type" means?}\xinyun{Is it clearer now?}\charles{Yes this is perfect. Thanks!}

In Figure~\ref{fig:res-flashfill}, we present the top-1 formula accuracies with different numbers of input data rows. We observe that even for spreadsheet formulas that only refer to cells in the same row, our models with tabular input encoders still perform better. 
\charles{What does it mean that RobustFill is given more than one row?}\xinyun{Not sure what is your question. Basically, RobustFill encodes each row as an IO sequence. Appendix~\ref{app:implementation-details} may make things clearer, but no space to discuss here.}\charles{Everything is OK now. This makes much more sense to me now, because we have clarified the above.}
In particular, with the increase of the number of input data rows, the accuracy of the RobustFill model does not show much improvement, while the accuracies of the other two models increase considerably, especially our full model. This demonstrates that our model could better utilize the available cell data context for prediction. Figure~\ref{fig:sheets-wrong-wo-data-ex} shows a formula that can be correctly predicted by our model when the full input context is given, but is wrongly predicted when the input only contains the header row and one data row. This example shows that understanding the cell data is especially important when the header is not informative enough. Notice that including only a few input rows or columns does not fit our encoder design well, since our BERT encoders simultaneously embed 3 data rows at a time, while the RobustFill model independently encodes each row by design. This could be the main reason why models with BERT-based encoders may perform worse than RobustFill when less than 3 data rows are presented. In addition, including headers still consistently provides a significant performance gain.~\rishabh{just to confirm our evaluation metric: in this setting, our notion of top-1 accuracy is that we produce syntactically the same formula as the ground truth, or are we executing the formula and we declare success if any formula that produces the correct target cell value}~\xinyun{The formula should be exactly the same. It is considered wrong even if the execution result is the same as the target cell value.}
\charles{I don't quite understand this. What does it mean
that our BERT encoders embed 3 data rows at a time,
for the models where there is only one data row
of context given to the model?}~\xinyun{Revised the sentence a bit, though not sure about your question. Appendix~\ref{app:implementation-details} may make things clearer, but no space to discuss here.}

\subsubsection{Results on Public Excel Spreadsheets}

Finally, we evaluate \ours{} on the Enron corpus~\footnote{The raw spreadsheet corpus is here:~\url{https://github.com/SheetJS/enron_xls}.}, which includes over 17K Excel Spreadsheets extracted from the Enron email corpus~\cite{klimt2004introducing,hermans2015enron}. We preprocess the Enron corpus in the same way as our Google Sheets corpus, and our final dataset includes 178K samples in the training set, 41K samples in the validation set, and 33K samples in the validation set. About $55\%$ formulas have sketch lengths of 2, $18\%$ formulas have sketch lengths of 3, $13\%$ formulas have sketch lengths of 4-5, $9\%$ formulas have sketch lengths of 6-7, and $5\%$ formulas have sketch lengths of at least 8. The formulas utilize 13 spreadsheet functions, and 4 general-purpose numerical operators (i.e.,~\texttt{+},~\texttt{-},~\texttt{*}, and~\texttt{/}). Compared to our Google Sheets corpus, the Enron dataset is smaller and the formulas include fewer types of spreadsheet functions, but it contain more formulas with long sketches. More details about the dataset are deferred to Appendix~\ref{app:data}.

On the Enron test set, \ours{} achieves $29.8\%$ top-1 accuracy, $41.8\%$ top-5 accuracy, and $48.5\%$ top-10 accuracy. These numbers are lower than the results on our Google Sheets corpus. When investigating into the model predictions, we observe that the main reason is due to the spreadsheet format difference. Specifically, because Enron spreadsheets are in Excel, while our data preprocessing pipeline is implemented for Google Sheets, we import Enron spreadsheets into Google Sheets for data preprocessing. Therefore, a larger proportion of table headers are not properly detected. However, when comparing to the prediction results without headers, as shown in Table~\ref{tab:res-full-exclude-headers}, the accuracies on the Enron test set are still better.
\section{Related Work}
\label{sec:work}

In this section, we present a high-level overview of the related work, and we defer a more in-depth discussion to Appendix~\ref{app:work}.
\emph{Program synthesis} has been a long-standing challenge, and various types of specifications have been discussed, including input-output examples~\citep{gulwani2012spreadsheet,balog2016deepcoder,bunel2018leveraging,bavishi2019autopandas,shin2018improving,chen2018execution}, natural language descriptions~\citep{gulwani2014nlyze,yu2018spider,yin2018learning,lin2018nl2bash,liang2018memory,wang2019rat}, and images~\citep{wu2017neural,liu2019learning,sun2018neural}. In particular, the FlashFill benchmark~\citep{gulwani2012spreadsheet} is the most related to our task, and their goal is to generate string transformation programs to manipulate the Excel spreadsheet data, given input-output examples as the specification. Various neural network approaches have been proposed for FlashFill~\citep{parisotto2016neuro,devlin2017robustfill,vijayakumar2018neural}. On the other hand, Nlyze~\citep{gulwani2014nlyze} translates natural language specifications to programs in an SQL-like DSL for spreadsheet data manipulation; and Autopandas~\citep{bavishi2019autopandas} synthesizes dataframe transformation functions implemented with the Python Pandas library, given input-output dataframe examples.
The spreadsheet formula prediction task in our work considers the semi-structured tabular spreadsheet context as the specification, rather than standardized input-output examples or natural language descriptions. Therefore, our formula specifications are more ambiguous and diverse. Furthermore, we show that including the header information is a key factor to improving the formula prediction performance.

In terms of the model input format, our spreadsheet formula prediction task is related to existing benchmarks on \emph{semantic parsing} over a tabular database~\citep{iyyer2017search,zhong2017seq2sql,yu2018spider}. There are two key differences between these tasks and ours. First, their program specification contains a natural language question, while our work predicts spreadsheet formulas based on the tabular context only. Therefore, our input specification is much more ambiguous. Meanwhile, our spreadsheet tables are typically less structured than the database tables. As shown in Figure~\ref{fig:sheets-ex}, spreadsheet tables do not necessarily satisfy a consistent row-based schema, and data cell values may be dependent on cells from other rows.

Our tabular context encoder is related to prior works on tabular BERT models, including TAPAS~\cite{herzig2020tapas}, TaBERT~\cite{yin2020tabert}, and Table-BERT~\cite{chen2019tabfact}. Our encoder design differs from these works in the following ways. First, these models are designed for question answering~\cite{herzig2020tapas,yin2020tabert} or fact verification~\cite{chen2019tabfact}, thus their inputs are the concatenation of a natural language question/statement and a table. In contrast, our model input only contains a spreadsheet table. Second, both TAPAS and Table-BERT require that the maximum table size is 512 tokens, which is not enough for our problem. SpreadsheetCoder encodes larger tabular input by tiling multiple rows/columns in multiple forward passes over BERT, and then doing the convolution to capture broader context. TaBERT independently embeds each table row with the question, then applies an attention mechanism over other tokens in the same column but different rows. This is similar to our row-based BERT without the row-wise convolution. As shown in Table~\ref{tab:res-full}, this alternative underperforms our full model.

Our spreadsheet formula prediction problem is also related to \emph{code completion} tasks~\citep{raychev2014code,li2018code,svyatkovskiy2019pythia,svyatkovskiy2020intellicode,svyatkovskoy2020fast}. Specifically, the goal of code completion tasks is to synthesize the subsequent program tokens given the code context, while we aim to generate the formula in the cell with the missing value to complete the spreadsheet. However, instead of providing a token sequence to represent the code context, our data context is a semi-structured table, where data values in different cells are connected in a two-dimensional space.
\vspace{-0.5em}
\section{Conclusion}
\label{sec:conc}
\vspace{-0.5em}

We presented the first technique to synthesize spreadsheet formulas given a tabular context, including both headers and cell values. In particular, we develop~\ours, a BERT-based model to capture the two-dimensional relational structure of the spreadsheet context, which are typically semi-structured tables. We demonstrate that incorporating the table headers significantly facilitates the prediction. Furthermore, modeling the dependency among cells of different rows and columns is important for generating formulas in real-world spreadsheets with diverse table structures. Compared to the rule-based system on Google Sheets, \ours{} assists 82\% more users in composing formulas.

There are a number of promising directions for future research about spreadsheet applications. First, developing a paradigm for pre-training on spreadsheet data could enable the encoder to be more specialized for spreadsheet applications. Second, we could infer more fine-grained knowledge of the table structure from the spreadsheet format information, such as colors and fonts, which could be utilized to develop more advanced encoder architectures. Finally, we could also extend our approach to support more spreadsheet applications, such as bug detection and clone detection.


\bibliography{ref}
\bibliographystyle{icml2021}

\clearpage
\appendix
\section{An Extended Discussion of Related Work}
\label{app:work}

Various neural network approaches have been proposed for the FlashFill benchmark~\citep{parisotto2016neuro,devlin2017robustfill,vijayakumar2018neural}. Specifically, both R3NN~\citep{parisotto2016neuro} and RobustFill~\citep{devlin2017robustfill} are purely statistical models, and RobustFill performs better. In a RobustFill model, each formula is executed on a single data row, thus each row is independently fed into a shared encoder. Afterwards, at each decoding step, a shared LSTM decoder generates a hidden state per data row, which are then fed into a max pooling layer. Finally, the pooled hidden state is fed into a fully-connected layer to predict the formula token. \maniatis{A little confused here. Doesn't the encoder generate a hidden per example, and then those are max-pooled to give an embedding for the input/output examples? Is the decoder LSTM involved in encoding?}~\xinyun{According to the RobustFill paper, they applied a late pooling (on page 5 of Section 4.3). Basically, the decoder takes one IO at a time, and it could attend to each token embedding generated by the encoder. Suppose we have 5 IO pairs, then we have 5 hidden states generated by the decoder per step. Afterwards, the max pooling is applied on these 5 hidden states, and the pooled vector is fed into the softmax layer. Therefore, max pooling is applied to the decoder hidden states, instead of the encoder. They discussed that this is one of the reasons why their approach is better than R3NN, an earlier neural model for the FlashFill benchmark.} On the other hand, in~\citep{vijayakumar2018neural}, they design a neural network to guide the deductive search performed by PROSE~\citep{polozov2015flashmeta}, a commercial framework for input-output program synthesis. A recent work proposes neural-guided bottom-up search for program synthesis from input-output examples, and they extend the domain-specific language of FlashFill to support more spreadsheet programs~\citep{odena2020bustle}. 

Besides formula prediction, some previous work has studied other applications related to spreadsheets, including smell detection~\citep{hermans2012detecting,cheung2016custodes,singh2017melford,azam2019spreadsheet}, clone detection~\citep{hermans2013data,dou2016detecting,zhang2020learning}, and structure extraction for spreadsheet tables~\citep{dong2019semantic,dong2019tablesense}. Our proposed encoder architecture could potentially be adapted for these spreadsheet tasks as well, and we leave it for future work.

\section{More Experimental Results}
\label{app:results}

\begin{figure}
    \centering
    \includegraphics[width=\linewidth]{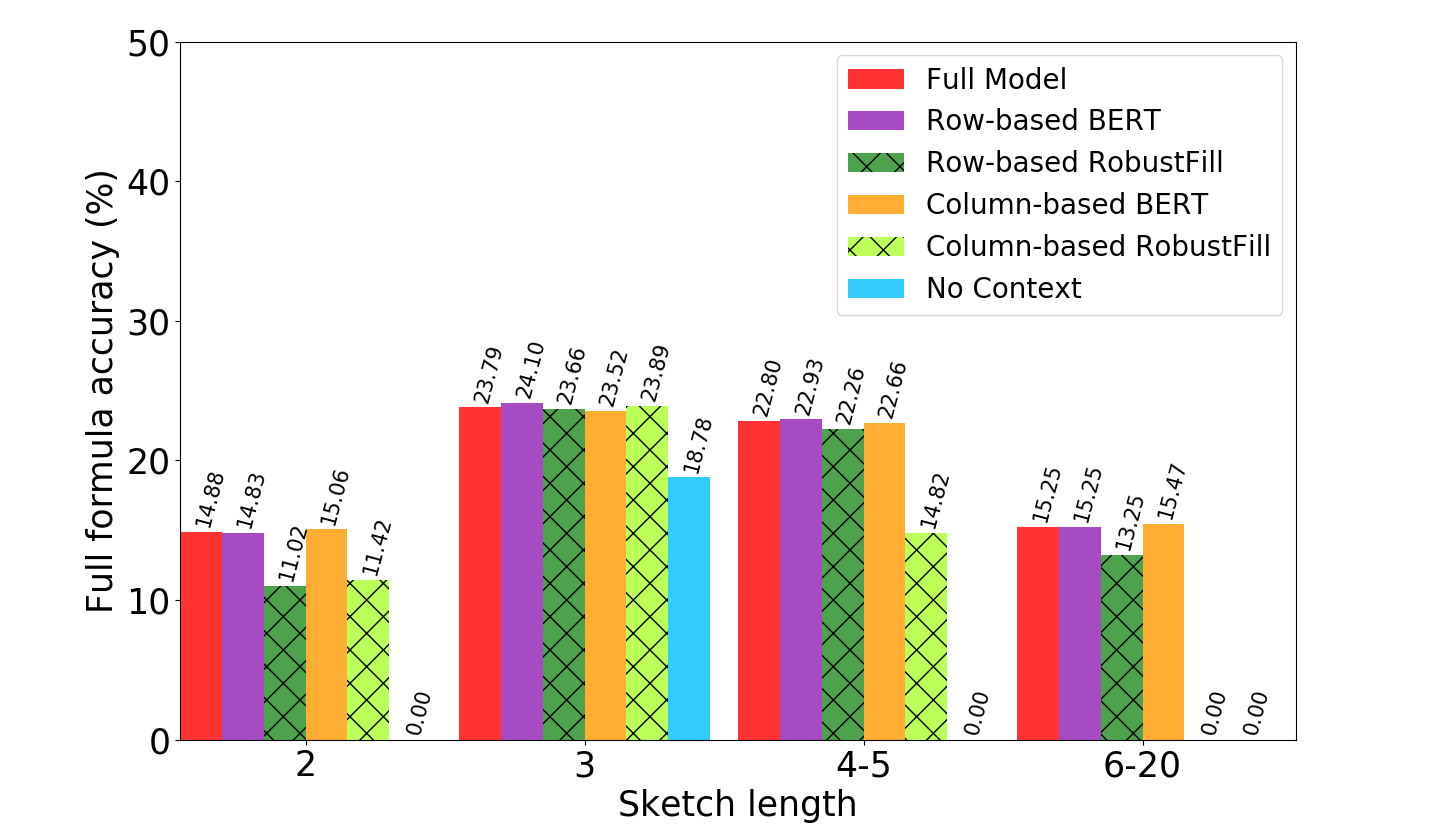}
    \captionof{figure}{Top-1 formula accuracies for different sketch lengths, excluding headers in the context.}
    \label{fig:res-sketch-length-exclude-headers}
\end{figure}

\begin{table}[t]
    \caption{Breakdown accuracies on the test set, excluding headers in the context.}
    \label{tab:res-breakdown-exclude-headers}
\begin{subtable}{\linewidth}
    \caption{Sketch accuracy.}
    \label{tab:res-sketch-exclude-headers}
    \centering
    \scalebox{0.85}{
    \begin{tabular}{rlccc}
    \toprule
    & \textbf{Approach} & \textbf{Top-1} & \textbf{Top-5} & \textbf{Top-10} \\
    \midrule
    &Full Model & 28.33\% & \textbf{62.55\%} & 72.89\% \\
    $-$ &Column-based BERT & 28.40\%  & 61.60\%  & \textbf{74.92\%}  \\
    $-$ &Row-based BERT & 27.71\% & 60.84\% & 73.43\% \\
    $-$ &Pretraining & \textbf{28.78\%} & 62.37\% & 74.61\% \\
    \midrule
    &Row-based RobustFill & 25.78\% & 42.66\% & 50.17\% \\
    &Column-based RobustFill & 26.15\% & 47.78\% & 57.72\% \\
    &No context & 25.19\% & 47.08\% & 52.70\% \\
    \bottomrule
    \end{tabular}}
\end{subtable}
\begin{subtable}{\linewidth}
    \caption{Range accuracy.}
    \label{tab:res-range-exclude-headers}
    \centering
    \scalebox{0.85}{
    \begin{tabular}{rlccc}
    \toprule
    & \textbf{Approach} & \textbf{Top-1} & \textbf{Top-5} & \textbf{Top-10} \\
    \midrule
    &Full Model & 22.60\% & 47.11\% & 53.84\% \\
    $-$ &Column-based BERT & 22.82\% & \textbf{47.76\%} & \textbf{54.98\%}  \\
    $-$ &Row-based BERT & 22.47\% & 46.14\% & 54.51\% \\
    $-$ &Pretraining & \textbf{23.48\%} & 47.27\% & 54.59\% \\
    \midrule
    &Row-based RobustFill & 21.01\% & 38.21\% & 43.89\% \\
    &Column-based RobustFill & 21.27\% & 37.80\% & 43.77\% \\
    &No context & 11.80\% & 25.54\% & 38.07\% \\
    \bottomrule
    \end{tabular}}
\end{subtable}
\end{table}

For the setting where the model input does not include headers, corresponding to Table~\ref{tab:res-full-exclude-headers} in Section~\ref{sec:exp-results-wo-headers}, we present the sketch and range accuracies in Table~\ref{tab:res-breakdown-exclude-headers}, and the breakdown accuracies on formulas of different sketch lengths in Figure~\ref{fig:res-sketch-length-exclude-headers}. We observe that the performance degradation is more severe for formulas of sketch lengths $2$--$3$.

\section{More Dataset Details}
\label{app:data}

Although in principle, our model could generate formulas using any operator in the spreadsheet language, some kinds of value references are impossible to predict from local context, thus we remove formulas with such values from our dataset. Specifically, we exclude formulas that use the \texttt{HYPERLINK} function with a literal URL, since those are merely "stylistic" formulas that perform no computation beyond presenting a URL as a clickable link. As discussed in Section~\ref{sec:setup}, we also filtered out formulas with cross-references from other tabs or spreadsheets. In total, the formulas filtered out after these two steps constitute around $40\%$ of all formulas. We further filtered out formulas with cell references farther than 10 rows or columns from the target cell in either direction, and formulas with absolute cell ranges. In this way, about $45\%$ of the original set of formulas are kept in our dataset.

Meanwhile, we observe that some spreadsheets may have tens of thousands of rows including the same formula, and including all of them in the dataset could bias our data distribution. Therefore, when multiple rows in the same spreadsheet table include the same formula in the same column, we keep the first 10 occurrences of such a formula, and create one data sample per formula. In this way, we extract around 846K formulas from 20M formulas before this filtering step, and we split them into 770K training samples, 42K for validation, and 34K for testing.

In total, around $100$ operators are covered in our output vocabulary. Among all spreadsheet formulas included in our filtered dataset, we list the 30 most commonly used spreadsheet functions and operators with their types~\footnote{The function types are based on the Google Sheets function list here:~\url{https://support.google.com/docs/table/25273?hl=en}.} as follows: \texttt{SUM} (Math), \texttt{+} (Operator, equivalent to~\texttt{ADD}), \texttt{-} (Operator, equivalent to~\texttt{MINUS}), \texttt{*} (Operator, equivalent to~\texttt{MULTIPLY}), \texttt{/} (Operator, equivalent to~\texttt{DIV}), \texttt{\&} (Operator, equivalent to~\texttt{CONCAT}), \texttt{AVERAGE} (Statistical), \texttt{LEN} (Text), \texttt{UPLUS} (Operator), \texttt{STDEV} (Statistical), \texttt{COUNTA} (Statistical), \texttt{MAX} (Statistical), \texttt{LEFT} (Text), \texttt{IFERROR} (Logical), \texttt{ABS} (Math), \texttt{MEDIAN} (Statistical), \texttt{UMINUS} (Operator), \texttt{CONCATENATE} (Text), \texttt{ROUND} (Math), \texttt{WEEKNUM} (Date), \texttt{AVERAGEA} (Statistical), \texttt{MIN} (Statistical), \texttt{COUNT} (Statistical), \texttt{TRIM} (Text), \texttt{COS} (Math), \texttt{SIN} (Math), \texttt{SINH} (Math), \texttt{TODAY} (Date), \texttt{IF} (Logical), \texttt{MONTH} (Date). We observe that most of these functions and operators are for mathematical calculation, statistical computation, and text manipulation. However, people also write conditional statements, and spreadsheet formulas for calculating the dates.

The spreadsheet functions and operators utilized in the Enron corpus are: \texttt{+} (Operator, equivalent to~\texttt{ADD}), \texttt{SUM} (Math), \texttt{-} (Operator, equivalent to~\texttt{MINUS}), \texttt{UPLUS} (Operator), \texttt{*} (Operator, equivalent to~\texttt{MULTIPLY}), \texttt{/} (Operator, equivalent to~\texttt{DIV}), \texttt{AVERAGE} (Statistical), \texttt{MIN} (Statistical), \texttt{MAX} (Statistical), \texttt{UMINUS} (Operator), \texttt{COUNT} (Statistical), \texttt{COUNTA} (Statistical), \texttt{ABS} (Math), \texttt{LN} (Math), \texttt{DAY} (Date), \texttt{WEEKDAY} (Date), and \texttt{STDEV} (Statistical).

\section{More Discussion of the FlashFill-like Setting}
\label{app:flashfill}
Following prior work on FlashFill~\citep{devlin2017robustfill,parisotto2016neuro,vijayakumar2018neural}, we evaluate model performance when different numbers of data rows are presented to the model as input. Specifically, when the input includes $1$--$11$ data rows, we grow the input from the target row upward. Our full data context includes $21$ data rows, with 10 rows above the target cell, 10 rows below the target cell, and 1 row where the target cell locates. Consistent with prior work, when we vary the number of input data rows during inference, we always evaluate the same model trained with the full data context including 21 data rows. Since RobustFill independently encodes each row, it supports variable number of input rows by design. For our models with the tabular input representation, we set the rows to be empty when they are out of the input scope, and apply a mask to indicate that the corresponding data values are invalid.

\section{Implementation Details}
\label{app:implementation-details}

\paragraph{Data preprocessing.} The content in each cell includes its data type and value, and we concatenate them as a token sequence. For example, \texttt{A2} in Figure~\ref{fig:sheets-ex-row} is represented as \texttt{num 0}. As discussed in Section~\ref{sec:encoder}, we concatenate all cell values in the same row as a token sequence, where values of different cells are separated by the ~\texttt{[SEP]} token. Each data row fed into the model includes $L=128$ tokens, and when the concatenated token sequence exceeds the length limit, we discard cells that are further away from the target cell. For column-wise representation, we produce token embeddings independently for each column-wise bundle $S_{cb}=[H_c, C_{3b-1}, C_{3b}, C_{3b+1}]$ for $b \in [-3, 3]$, where $C_i$ is a token sequence produced by concatenating all tokens of the cells in column $C_i$.

\paragraph{Output vocabulary construction.} To construct the output formula token vocabulary, we filtered out tokens that appear less than 10 times in the training set, so that the vocabulary contains 462 tokens, out of 2625 tokens before filtering. In total, around a hundred operators are covered in our output vocabulary, including 82 spreadsheet-specific functions, and other general-purpose numerical operators (e.g.,~\texttt{+},~\texttt{-}).

\paragraph{Hyper-parameters.} The formula decoder is a 1-layer LSTM with the hidden size of 512. We train the model with the Adam optimizer, with an initial learning rate of 5e-5. We train models for 200K minibatch updates, with a batch size 64. We set the dropout rate to be 0.1 for training. The norm for gradient clipping is 1.0.

\end{document}